\documentclass[conference]{IEEEtran}
\IEEEoverridecommandlockouts
\usepackage{cite}
\usepackage{amsmath,amssymb,amsfonts}
\usepackage{algorithmic}
\usepackage{graphicx}
\usepackage{textcomp}
\usepackage{xcolor}

\usepackage{amsmath}

\usepackage{listings}
\usepackage[most]{tcolorbox}
\usepackage{inconsolata}

\newtcblisting[auto counter]{mylisting}[2][]{sharp corners, 
    fonttitle=\bfseries, colframe=gray, listing only, 
    listing options={basicstyle=\ttfamily,language=java}, 
    title=Code-block \thetcbcounter: #2, #1}

\def\BibTeX{{\rm B\kern-.05em{\sc i\kern-.025em b}\kern-.08em
    T\kern-.1667em\lower.7ex\hbox{E}\kern-.125emX}}
\begin{document}

\IEEEoverridecommandlockouts
\IEEEpubid{\makebox[\columnwidth]{978-1-7281-2747-7/20/\$31.00~\copyright2020 IEEE \hfill} \hspace{\columnsep}\makebox[\columnwidth]{ }}

\title{Multi-dimensional Skyline Query to Find Best Shopping Mall for Customers
}

\IEEEpubidadjcol

\author{\IEEEauthorblockN{Md Amiruzzaman}
\IEEEauthorblockA{
Kent State University\\
Kent, OH, USA \\
mamiruzz@kent.edu}
\and
\IEEEauthorblockN{Suphanut Jamonnak}
\IEEEauthorblockA{
Kent State University\\
Kent, OH, USA \\
sjamonna@kent.edu}
}


\maketitle

\begin{abstract}
This paper presents a new application for multi-dimensional Skyline query. The idea presented in this paper can be used to find best shopping malls based on users requirements. A web-based application was used to simulate the problem and proposed solution. Also, a mathematical definition was developed to define the problem and show how multi-dimensional Skyline query can be used to solve complex problems, such as, finding shopping malls using multiple different criteria. The idea of this paper can be used in other fields, where different criteria should be considered.\\
\end{abstract}

\section{INTRODUCTION}
In market research finding best shopping mall for customers or consumers is an important task \cite{jarvenpaa1996consumer, wakefield1998excitement}. Especially, the problem becomes more complicated as consumers behavior and choices may differ \cite{childerhouse2000engineering}. In this paper the word consumer and customer are used interchangeably. There are several factors that may play a vital role in consumers’ choice for a shopping mall. For example, distance of the shopping mall from customers’ current location, price of goods, parking spaces in the shopping mall. Of course, free parking spaces may attract most customer; especially in big cities \cite{hasker2014free}.\\

Jarvenpaa and Todd \cite{jarvenpaa1996consumer} mentioned that web-based apps are another attraction for modern customers. Often, customers like to browse items (i.e., good), and check their prices, and availability of the product before they finally arrive at the shopping mall or store.  Wakefield and Baker \cite{wakefield1998excitement} reported that often customers may go to a shopping mall that is less crowded. Because, chance of finding an item in a less popular shopping malls or less crowded shopping malls is higher \cite{mittelstaedt1990shopping}. This phenomenon indicates that consumer selecting a shopping mall depends of on many uncertain or probabilistic factors \cite{can2016case}, so, ahead of time it is not easy to compute which shopping mall is best fit for customer ahead of time. However, with the help of uncertain or probabilistic data management, it is possible to compute customer choices with some certainty.\\

Whereas it is important to find factors that satisfies customer and provides compatible prices \cite{hunneman2017moderating}. It is important to develop a web-based app that helps customer to find one of the best shopping malls, and help business owners to understand what factors matter the most to customers. The basic Skyline query is described by Pei, Jiang, Lin and Yuan \cite{pei2007probabilistic}. In their work they described the strength of Skyline in multi-factor decision making. Pei, Jiang, Lin and Yuan \cite{pei2007probabilistic} described the use of Skyline query to find one of the best National Basketball Association (NBA) player for a game, considering number of rebounds and number of assists. The problem they described can be plotted in a two-dimensional space. Where one dimension is number of rebounds and another dimension is number of assists.\\

Similarly, finding a cheaper hotel near a beach or attraction can be solved using traditional Skyline query \cite{kossmann2002shooting, borzsony2001skyline}. Where managing both parameters can be a challenge. However, Skyline query can retrieve a result that can be helpful for customers \cite{jin2007multi, dellis2007efficient}. This type of query can be helpful in location and attraction-based business planning and predicting the success of the business \cite{zhu2009efficient}. For example, if a system should considers users preferences, such as free parking spaces, cheaper products, access to restaurant, restrooms, access to car mechanic shops, etc., then finding a suitable shopping mall can be a challenging task. This problem seems to be a multi-dimensional customers choice, where choices are probabilistic in nature. Therefore, this problem can be solved using a modified Skyline query or multi-dimensional Skyline query.\\

This paper presents a web-based system focusing on customers’ stratification. The web-based tool is was developed with the help of multi-dimensional Skyline query. The system provides ranks of shopping malls based on customers’ preferences and helps customers to find one of the best shopping malls. 

\section{Existing methods}\label{lit}
There are a few studies that very closely related to this project; all these are describing how to compute the Skyline probabilities on uncertain data \cite{dheenadayalan2014premonition}. This literature review focused on basic methods and their contributions to compute the Skyline probabilities. For example, Pei, Jiang, Lin, and Yuan \cite{pei2007probabilistic} presented a probabilistic Skylines on uncertain data, it addresses two major challenges about Skyline analysis and computation on uncertain data. The first challenge is modeling Skylines on uncertain data and proposed the notion of probabilistic Skyline and introduces the probabilistic nature of uncertain objects into the Skyline analysis. The probability of an object being in the Skyline is the probability that the object is not dominated by any other objects. The second challenge is efficient computation of probabilistic Skylines; the authors developed two algorithms to tackle this problem.\\

First, the bottom-up algorithm computes the Skyline probabilities of some selected instances of uncertain objects and uses those instances to prune other instances and uncertain objects effectively \cite{borzsony2001skyline, pei2007probabilistic}. Second, the top-down algorithm recursively partitions the instances of uncertain objects into subsets, and prunes subsets and objects aggressively \cite{pei2007probabilistic}. In a study, Atallah and Qi \cite{atallah2009computing},  presented a work based on Pei, Jiang, Lin, and Yuan’s \cite{pei2007probabilistic} work and provided the first Sub-Quadratic algorithm to compute all Skyline probabilities and some new probabilistic Skyline analysis, their work can deal with more general uncertain data model. Khalefa, Mokbel and Levandoski \cite{khalefa2010skyline} proposed an efficient framework that supports Skyline queries for uncertain data represented as a continuous range.

The rest of this paper is organized as follows: The section \ref{prob} presents the mathematical definition of the problem. Section \ref{method} describes data collection, system design, interface used to simulate the problem and solution. The experimental results are presented in \ref{exprimental} section. Finally, the section \ref{conclusion} concludes the study and provides future research directions.

\section{Problem Definition}\label{prob}

Let, $U = \{u_1, u_2, u_3,\cdots, u_n\}$ are users/customer who shops in different shopping malls, $S = \{s_1, s_2, s_3,\cdots, s_n\}$. However, users have preference which can be represented as keyword $K = \{k_1, k_2, k_3, \cdots, k_n\}$. The distance of each shopping malls are $D = \{d_1, d_2, d_3, \cdots, d_n\}$, and price of products $C = \{c_1, c_2, c_3,\cdots, c_n\}$, however, we may consider total cost or sum of products, i.e.,  $\sum_{i=1}^{n}c_i$. Based on previous visits of each shop, probability to pick a shop can be denoted as, $P = \{p_1, p_2, p_3,\cdots, p_n\}$.\\ 

\begin{figure}
\begin{center}
\includegraphics[width=0.4\textwidth]{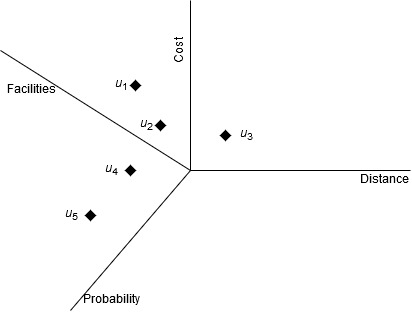}
\caption{Multi-dimensional Skyline problem. Different dimensions are shown using different lines, and $u_1, u_2, \cdots, u_5$ are different users.}\label{skyline}
\end{center}
\end{figure}

Note that, different shops malls may sale different types of goods, $G = \{g_1, g_2, g_3,\cdots, g_n\}$, and facilities (e.g., restaurant, kids zone, bar, etc.) in each shopping mall may vary as well, $F = \{f_1, f_2, f_3, \cdots, f_n\}$. 
If we are interested to know about a particular user (i.e., query user, $u_q$), then this problem can be represented as a multi-dimensional Skyline problem. As such, shorter distance, lower cost, more facilities, higher variety of goods are desirable.  Also, for the simplicity of the problem, we will consider higher probability or mostly visited shopping malls first (see Figure \ref{skyline}). 

\section{Methodology}\label{method}
\subsection{Data Collection and Pre-processing}
In order to develop our system. First, the research data was gathered. That means the shopping mall data was collected from the shopping mall directory website (https://shoppingcenters.com/). For the simulation, the data set only focused on Cleveland and Akron areas. These are two popular cities from the State of Ohio, United states. In addition, detail reports from each shopping malls were manually saved. Thus, a total of 90 shopping mall reports used for the simulation and experiment. The report shown that several attributes can be considered as multi-dimension in Skyline-Query presented on this paper. Table \ref{tbl:attributes} shows all selected attributes that have been used to apply our algorithms and techniques.\\

\begin{table*}[]
\caption{\label{tbl:attributes}Multi-Dimension and Attributes.}
\begin{tabular}{p{40mm}p{20mm}p{107mm}}
\hline\hline
Attributes               & Type    & Definition                                                               \\\hline\hline
Store number             & Integer & Indicate number of the stores both inside and outside of a shopping mall \\
Parking space            & Integer & Number of available parking spaces in a shopping mall                    \\
Food court               & Boolean & Whether or not Shopping provide food court                               \\
Average household income & Integer & Average of income around shopping mall area                              \\
Facilities               & Integer & Facilities type categorized by stores name\\\hline                              
\end{tabular}
\end{table*}

The research data was pre-processed using the mentioned attributes (see Table \ref{tbl:attributes}) and divided into two phases, as follows:
\begin{enumerate}
    \item Generate a geo-location (Latitude and Longitude) for each shopping mall
    \item Calculate the total sum for each facility provided by each shopping mall, which including (Anchor, Services, Miscellaneous, Hi-Tech, Restaurants, Specialty, Barbers and Beauty, Women’s wear, Men’s wear, Unisex and Family Clothing, Shoes, Children Apparel, Gifts Cards and Books, Jewelry, Entertainment).
\end{enumerate}{}
These total sums have been stored in an array of integer. For instance, facilities attribute can be illustrated as $[ 3, 3, 1, 2, 6, 0, 7, \cdots]$.\\

Lastly, manually computed this process by filling out in the spreadsheet; in order to create a data in Comma-Separated Values (.csv) format. Thus, data was imported to the PostgreSQL database \cite{postgresql1996postgresql}. Table \ref{tbl:schema}. Illustrates an example rows and columns in PostgreSQL schema \cite{schema2001data}.

\subsection{System design}
\begin{figure*}
\begin{center}
\includegraphics[width=0.9\textwidth]{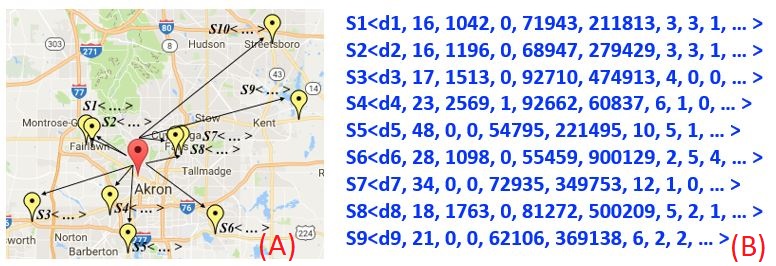}
\caption{Example of 10 shopping malls in a set of tuples. (A) is the map view. Yellow markers are shopping malls and Red marker is the user's current location. (B) is the tuple view of each shopping mall. }\label{example10shop}
\end{center}
\end{figure*}

In the proposed system, each shopping mall was formed as a set of tuple $S_i$, in which each tuple consists of $S_i<$\emph{Distance, Stores Number, Parking Space, Food Court, Average Household Income, Population, Facilities}$>$.\\
Figures \ref{example10shop} shows 10 tuples visualized on a map in yellow marker. The red marker is the user’s selected location, and arrow illustrate a set of distance $D_i$, which used to calculate a distance from user’s selected location to each shopping mall.\\ 

In addition, the distance from user's current location was added to the tuple as it was pre-computed and calculated based on user’s selected location. Also, driving distance matrix was constructed using a Google Direction Services API \cite{le2015application}, which helped to compute a driving distance from the starting point (user’s selected location) to the ending point (shopping mall location).\\

\begin{table*}[]
\caption{\label{tbl:schema}Table in PostgreSQL schema.}
\begin{tabular}{cccccccccc}
\hline\hline
Mall & Code & Lat & Lng & Store Number & Parking space & Food court & Average household income & Population & Facilities       \\\hline\hline
$S_1$ & OH1  &  41.502744 & -81.502225  & 16           & 1042          & 0          & 71,943                   & 211813     & {[}3,3,1,...{]}  \\
$S_2$ & OH2  &  41.463094  & -81.476332  & 16           & 1196          & 0          & 68,947                    & 279429     & {[}3,3,1,...{]}  \\
$S_3$ & OH3  &  41.499291   &  -81.492427   & 17           & 1513          & 0          & 92,710                    & 474913     & {[}4,0,0,...{]}  \\
$S_4$ & OH4  &  41.381915   & -81.742649    & 23           & 2569          & 1          & 92,662                    & 60837      & {[}6,1,0,...{]}  \\
$S_5$ & OH5  &  41.458837   & -81.951638    & 48           & 0             & 0          & 54,795                    & 221495     & {[}10,5,1,...{]}\\\hline
\end{tabular}
\end{table*}

Thus, The proposed solution and an application of the problem was simulated using a web-based application framework. The web application was developed using HTML5, JavaScript, CSS as a front-end, and back-end was developed using PHP and external libraries such as Google Maps API and Google Direction Services. The PHP language was used as a back-end scripting languages to communicate between the web application and PostgreSQL database.\\

\begin{figure}
\begin{center}
\includegraphics[width=0.5\textwidth]{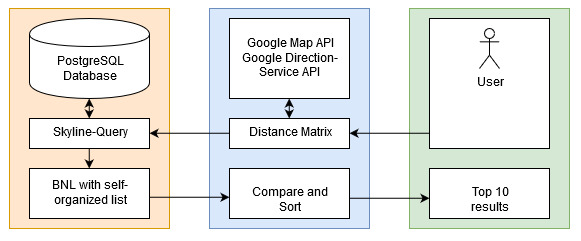}
\caption{Framework for the proposed work. The Block diagram shows step-by-step process for data acquisition and processing. Finally, the system provides 10 different results in ascending order.}\label{framework}
\end{center}
\end{figure}

\subsection{Proposed Algorithms and techniques}
The three most common algorithms for skyline queries are Block-Nested-Loop (BNL), Divide and Conquer (DAC), and Branch and Bound Skyline (BBS). Following paragraphs describes each algorithm briefly.\\

\begin{enumerate}
    \item BNL algorithms \cite{atallah2009computing, keles2019skyline} are the most intuitive approach to processing Skyline queries; wherein each data point is compared with other data points in the database to check whether it is dominated by other data points. If not, then the data point is listed as a candidate Skyline data point and may become the final result of the Skyline query. If the data point is dominated by another data point, then it cannot be the final result of the Skyline query and is eliminated. Once all comparisons are completed, the remaining candidate points form the final Skyline query results. 
    \item DAC algorithms \cite{atallah2009computing, wei2019assigning} first divide the data in a database into smaller groups, process the groups, integrate the skyline data points from all of the groups, and then perform another skyline query to obtain the final results. 
    \item BBS algorithms \cite{atallah2009computing, bouderar2019solving} are currently the most frequently used because they index the data in order to reduce the number of data points that need to be checked. In contrast, BNL algorithms check all of the data points, and DAC algorithms check most of the data points. 
\end{enumerate}

In this paper, the proposed solution was implemented using two approaches, such as, 

\begin{enumerate}
    \item First, the Skyline-query was implemented using a regular SQL command
    \item Second, the Blocked-Nested-Loop (BNL) was implemented with some revised version, which is referred as sort-filter-skyline \cite{du2019two}.
\end{enumerate}

\subsection{Skyline-operator}
Börzsönyi et al. \cite{borzsony2001skyline} purposed a Skyline operation as an extension of SQL query. Its original example of application of Skyline operator involves selecting a hotel for a holiday \cite{magnani2013skyview}. Suppose, a user wants the hotel to be both cheap and close to the beach. However, hotels that are close to the beach may also be expensive. In this case, the Skyline operator would only present those hotels that are not worse than any other hotel in both price and distance to the beach. Similar to our problem, we would like to select a shopping mall base on shorter driving distance, more stores, more available parking space, lower household income and lower in population. As you seen, our problem requires several dimensions to be considered, which required more dimensions than the hotel problem. As a result, code-block\#1 transform a purposed Skyline operator related to our shopping mall problem.\\

\begin{mylisting}[colback=white]{Transform Skyline operator}
SELECT * FROM global
SKYLINE OF 	Distance MIN, 
    Stores Number MAX,
    Parking Space MAX,
    Household Income MIN,
    Population MIN, ... 
\end{mylisting}

We ranked shopping mall by the nearest Distance, which dynamically generated based on the user’s selected location. We used Google Driving Distance Matrix \cite{DistanceMatrixAPI} in order to generate a driving distance from user’s selected location to each shopping mall in our database. We also select a shopping mall which has more stores and more parking space than the other. Because we believe that customers are prefer more buying options and more comfortable to park their car. Our application also generates much more dimension base on user’s selected preferences such as Clothing, Restaurants, and Services. These will dynamically add to our SQL query as much as they prefer. Thus, we purpose a Skyline query as a regular SQL command, which implemented using PHP and PostgreSQL database to generate a several shopping malls which not worse than the other. Code-block\#2 illustrates our purpose SQL command similar to Skyline operator.

\begin{mylisting}[colback=white]{Our purpose SQL command}
SELECT *
FROM Shopping Mall S
WHERE NOT EXISTS (
SELECT * FROM Shopping Mall S1
    AND S1.Distance <= S.Distance
    AND S1.StoresNum >= S.StoresNum
    AND S1.ParkingSpace 
        >= S.ParkingSpace
    AND (S1.Distance < S.Distance OR
    	S1.StoresNum > S.StoresNum OR
    	S1.ParkingSpace > 
\end{mylisting}

\subsection{Block-Nested-Loop (BNL)}
Due to several dimensions and attributes, the proposed Skyline query would generate redundant shopping mall and ambiguous results. As a result, a revised version of the BNL algorithms referred to as the sort-filter-skyline (SFS) algorithm was implemented. This algorithm first calculates the sum of all dimensions for each data point and then ranks the data points in ascending order according to magnitude. As a result, we need only to check whether lower ranking data points are dominated by higher ranking data points. There is no need to check whether higher ranking data points are dominated by lower ranking data points because data points with smaller sums are never be dominated by those with greater sums \cite{chiu2015finding}. This approach greatly increases the overall processing speed. Code-block\#3 illustrates class Block-Nested-Loop (BNL) join.

\begin{mylisting}[colback=white]{Block-Nested-Loop (BNL) join}
For each block Br of r do
  For each block Bs of s do
    For each tuple tr in Br do
      For each tuple ts in Bs do
      	Check if (tr,ts) join 
      	  add tr = ts to the result
      End
    End
  End
End
\end{mylisting}

The proposed application generates top shopping mall result in two parts. Both results comes from SQL query and BNL algorithm. Next, it  generates a final result using matching algorithm to compare a similar shopping mall, which is generated from both SQL query and BNL algorithm. Thus, it generates a final shopping mall result along with its probability values illustrated in Table \ref{tbl:prob}.

\begin{table*}[]
\caption{\label{tbl:prob}Final result with probability values.}
\begin{tabular}{cccccccc}
\hline\hline
Mall & Store number & Parking space & Food court & Average household income & Population & Facilities              & Probability \\
\hline\hline
$S_1$ & 16           & 1042          & 0          & 71,943                   & 21813      & {[}3, 3, 1, ...{]} & 0.50        \\
$S_2$ & 16           & 1196          & 0          & 68,947                   & 279429     & {[}3,3,1,...{]}    & 0.43        \\
$S_3$ & 17           & 1513          & 0          & 92,710                   & 474913     & {[}4,0,0,...{]}     & 0.61        \\
$S_4$ & 23           & 2569          & 1          & 92,662                   & 60837      & {[}6,1,0,...{]}         & 0.70        \\
$S_5$ & 48           & 0             & 0          & 54,795                   & 221495     & {[}10,5,1,...{]}        & 0.20\\\hline       
\end{tabular}
\end{table*}

\subsection{Interface}
\begin{figure}
\begin{center}
\includegraphics[width=0.5\textwidth]{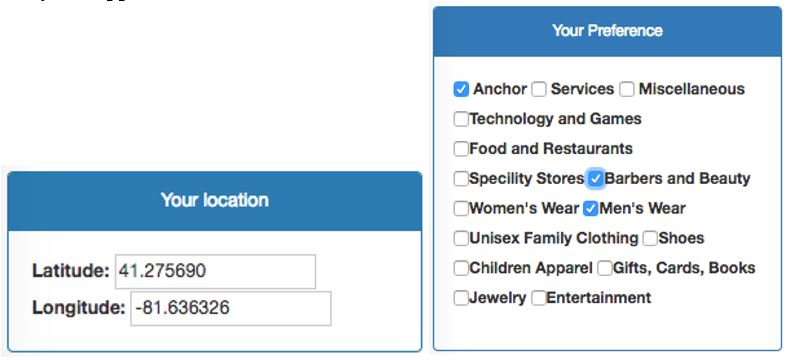}
\caption{User’s dynamic input. (left) user's location using latitude and longitude, (right) user's preferences.}\label{userDynamicInput}
\end{center}
\end{figure}

\begin{figure*}
\begin{center}
\includegraphics[width=\textwidth]{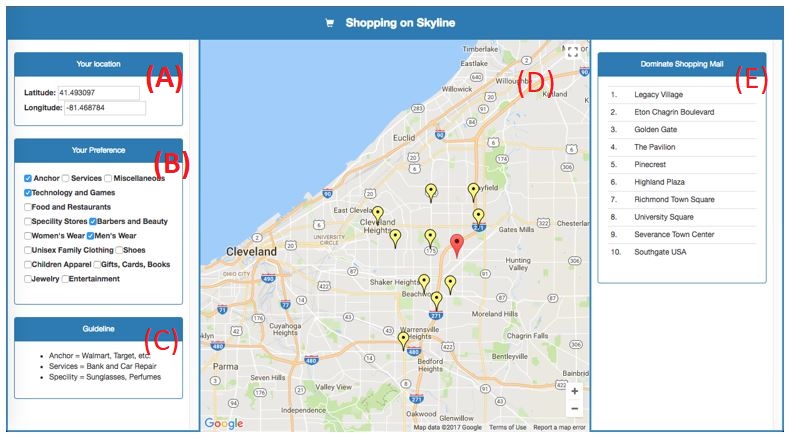}
\caption{The proposed web application and shopping mall results. (A) user's current location, (B) user's preferences, (C) Guidelines, (D) map view to provide visual location of user and shopping malls, (E) final result provided by the proposed Skyline algorithm.}\label{application}
\end{center}
\end{figure*}

The user’s input interfaces are shown in Figure \ref{userDynamicInput}, where users can drag a red marker to specify their preferred location or their current location. This will automatically be updated in the “Your Location” panel (see Fig. \ref{userDynamicInput}). In addition, users can select which facilities or categories of product their preferred such as Anchor, Services, Technology, Restaurants, and Specialty Stores. This can be performed by checking a check box in a “Your Preference” panel. After users specify their location and their preference, our application will generate a top shopping mall result with a yellow marker drawn on a map. Figure \ref{userDynamicInput} shows a dynamic user’s input. In addition, Figure \ref{application} illustrates an example result generated by the proposed web application.

\section{Experimental Evaluation}\label{exprimental}
An experiment was conducted with initial data with 90 records. The experiment consists of Cuyahoga, Summit, Lorain, Stark, Medina, Lake, Wayne, and Portage county. All these counties are from Northeast Ohio. The proposed method (i.e., algorithm) and dimensionality is completely new; no existing algorithms found that can be compared with. Therefore, comparisons were not made. However, it is the intention to continue this work with a large-scale database and more parameters. 
Experimental data has shown 100\% success in query. Maybe, this was because of the nature of the data and size, these results were achieved. At a later time, large scale data will be considered, and results will be evaluated again. 

\section{Conclusion and Future work}\label{conclusion}
During the literature review, it was noticed that there are different types of skyline-queries, however, among all only SQL command-based algorithm and BNL methods was considered and modified to accomplish the proposed algorithm. One of the drawbacks of SQL command based and BNL algorithm is that they both are cumbersome in nature, expensive to evaluate, and provided a huge query result set. So, both BNL and SQL command-based algorithms can be improved to provide a smaller result set and also reduce the complexity. Also, user study will be conducted to improve the reliability and usability of the proposed tool. User interface will be modified further to satisfy users need and color choice.

\bibliographystyle{IEEEtran}
\bibliography{IEEEabrv, bibfile}
\end{document}